\documentclass[11pt,a4paper]{article}
\usepackage[utf8]{inputenc}
\usepackage[english]{babel}
\usepackage{amsmath}
\usepackage{amsfonts}
\usepackage{amssymb}
\usepackage{makeidx}
\usepackage{graphicx}
\usepackage{authblk}
\usepackage[left=2cm,right=2cm,top=2cm,bottom=2cm]{geometry}
\author{Sebastian Jaksch}
\affil{J\"ulich Centre for Neutron Science (JNCS) at Heinz Maier-Leibnitz Zentrum (MLZ), Forschungszentrum J\"ulich GmbH, Lichtenbergstr. 1, 85748 Garching, Germany}
\title{Considerations about Chopper Configuration at a time-of-flight SANS Instrument at a Spallation Source}

\begin{document}
\maketitle

\begin{abstract}
In any neutron scattering experiment the measurement of the position of the scattered neutrons and their respective velocities is necessary. In order to do so, a position sensitive detector as well as a way to determine the velocities is needed. Measuring the velocities can either be done by using only a single wavelength and therefore velocity or by creating pulses, where the start and end time of each pulse is known and registering the time of arrival at the detector, which is the case we want to consider here. This pulse shaping process in neutron scattering instruments is usually done by using a configuration of several choppers. This set of choppers is then used to define both the beginning and the end of the pulse. Additionally there is of course also a selection in phase space determining the final resolution that can be achieved by the instrument. Taking into account the special requirements of a specific instrument, here a small-angle neutron scattering instrument, creates an additional set of restrictions that have to be taken into account.
In this manuscript a chopper configuration for two possible settings, namely a maximum flux and a high-resolution mode will be presented.
\end{abstract}

\section*{Introduction}
In neutron scattering the determination of position and velocity of the scattered neutrons is essential. While at continuous neutron sources often monochromators are used for the latter, this is not optimal at pulsed spallation sources. The reduction in flux by using a monochromator is considerable, which is why there are also time-of-flight (TOF) instruments at continuous sources \cite{Dewhurst}. Even the resolution in wavelength is inverse to the transmitted flux resulting the in the unsavory choice between high flux or high resolution. This can be alleviated using a chopper setup and a time resolving detector. At a pulsed spallation source also the \textit{natural} frequency of the source can be exploited to gain maximum flux at each given configuration. 

For all practical applications during these considerations SKADI \cite{jaksch2014} based at the ESS \cite{HALLWILTON2013} will be used as a model, however most of the ideas presented here are generally applicable. After the description as given in the original proposal there have been several changes in the layout of the facility. The only change with an impact on the considerations that are presented here is the introduction of a bunker wall between 11.5 and 15.0 m from the source, creating a dead area where not components can be placed. 

\begin{figure}[ht]
\begin{center}
\includegraphics[width=0.8\textwidth]{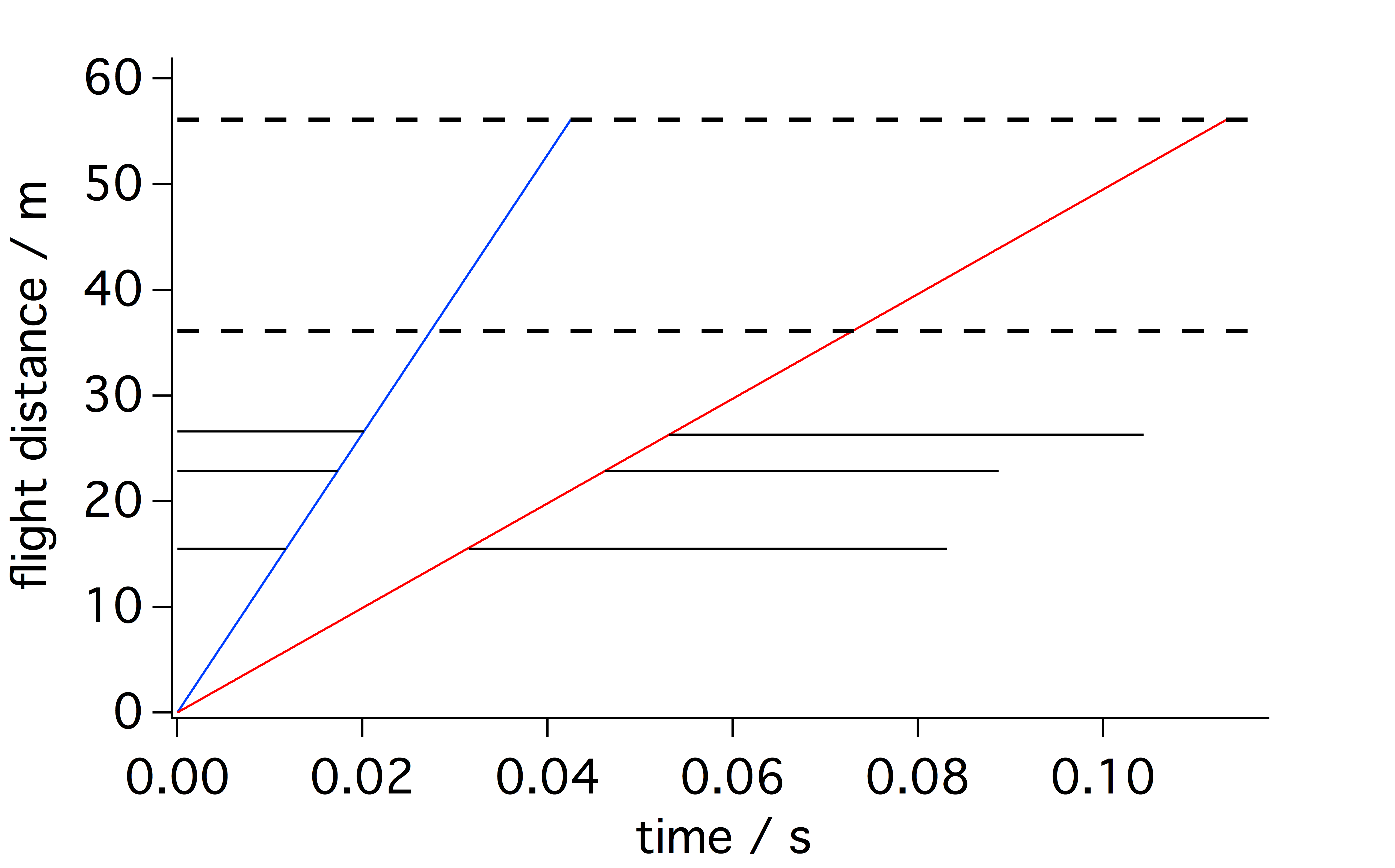}
\end{center}
\caption{Setup with all choppers as described in table \ref{tab:description}. The choppers at 26.3 and 26.6 m are running at 14 Hz in this case. For the high resolution setup they rotate at 15 times that frequency. Black dashed lines show the positions of the front and rear detector, blue line indicates the path of the 3\AA neutrons, red line indicates the path of the 8\AA neutrons.}
\label{fig:highfluxsetup}
\end{figure}

The resulting chopper setup, for which the optimization is detailed in this manuscript comprises of the choppers as shown in table \ref{tab:description} and fig.\ref{fig:highfluxsetup}.

\begin{table}
\begin{center}
\begin{tabular}{c|c|c|c|c}

Chopper description &$x_{chopper}$/m & frequency / Hz & Opening time / s & Opening in degree\\ 
\hline 
Frame Overlap  & 15.5 & 14 & 0.01957 & 98.64\\ 
\hline 
Higher Order Suppression & 22.85 & 14 & 0.02885 & 145.41\\ 
\hline
High Resolution 1 & 26.3 &$N\times 14$ & 0.03321 & 167.37\\
\hline
High Resolution 2 & 26.6 &$N\times 14$ & 0.03359 & 169.27\\

\end{tabular}
\end{center}
\caption{Positions, speeds and openings of the optimized chopper setup for SKADI. The opening time is for the standard mode optimized for high flux with $N=1$. In order to have a smaller opening, smaller openings at lower speeds can be used.}
\label{tab:description} 
\end{table}

This setup both provides a well defined neutron spectrum as well as the possibility to choose between high-flux and high-resolution experiments.

\section{Flux optimized configuration}
Most, if not all, neutron scattering instruments aim for a high flux of neutrons at the sample position. This is however counter current to a good resolution as due to the Luiville theorem phase space cannot be compressed, thus only the needed neutrons are selected, but no neutrons that were unsuitable for scattering can be converted into useful neutrons. To maximize the flux in this context means minimizing the amount of neutrons that are cut off in the instrument to achieve a good resolution both in space and time.

In the case of a SANS machine at a pulsed source this means that along the flight path the neutron flux is restricted angularly and in time. Here the consideration is focused on the time restriction.

In a single pulse the full wavelength band is emitted and has time until the next pulse arrives at the detector. All neutrons that take longer to travel that distance will be seen in the wrong pulse and therefore create a frame overlap. Thus the time differential between fast and slow neutrons in one pulse impinging on the detector $\Delta t$ depends on the length of the instrument and the wavelength of the neutrons in the pulse as follows

\begin{equation}
\Delta t = \frac{D\cdot m_n}{h} \left(\lambda _{slow}- \lambda_{fast}\right).
\label{eq:DeltaLambda}
\end{equation}

Here $D$ is the distance between source and detector, $m_n$ the mass of the neutron, $h$ the Planck constant, and $\lambda_{slow}$ and $\lambda_{fast}$ the wavelength of the slow and fast neutrons respectively. For future use in this manuscript $C=h/m_n = 3.96\cdot 10^{-7}\,\frac{\mbox{m}^2}{\mbox{s}}$ will be used as a constant. Knowing the repeat time of the source $T$ this renders the condition $\Delta t \leq T$ to avoid frame overlap. One limit of the wavelength band can thus be chosen freely, which usually means that the lower end of the wavelength band is fixed into a region with high flux from the source. It is immediately visible that a short instrument would result in a small $\Delta t$ and thus to a high flux. In case of SKADI this is chosen to be $\lambda_{fast}= 3\,$\AA\, at $f=\frac{1}{T}=14\,$Hz at a length of $D=56.11\,$m. Using eq.\ref{eq:DeltaLambda} this renders

\begin{equation}
\lambda_{slow} \leq T \frac{C}{D} + \lambda_{fast} = 5.04 \mbox{\AA}+ \lambda_{fast} = 8.04 \mbox{\AA}.
\end{equation}

\begin{figure}[ht]
\begin{center}
\includegraphics[width=0.8\textwidth]{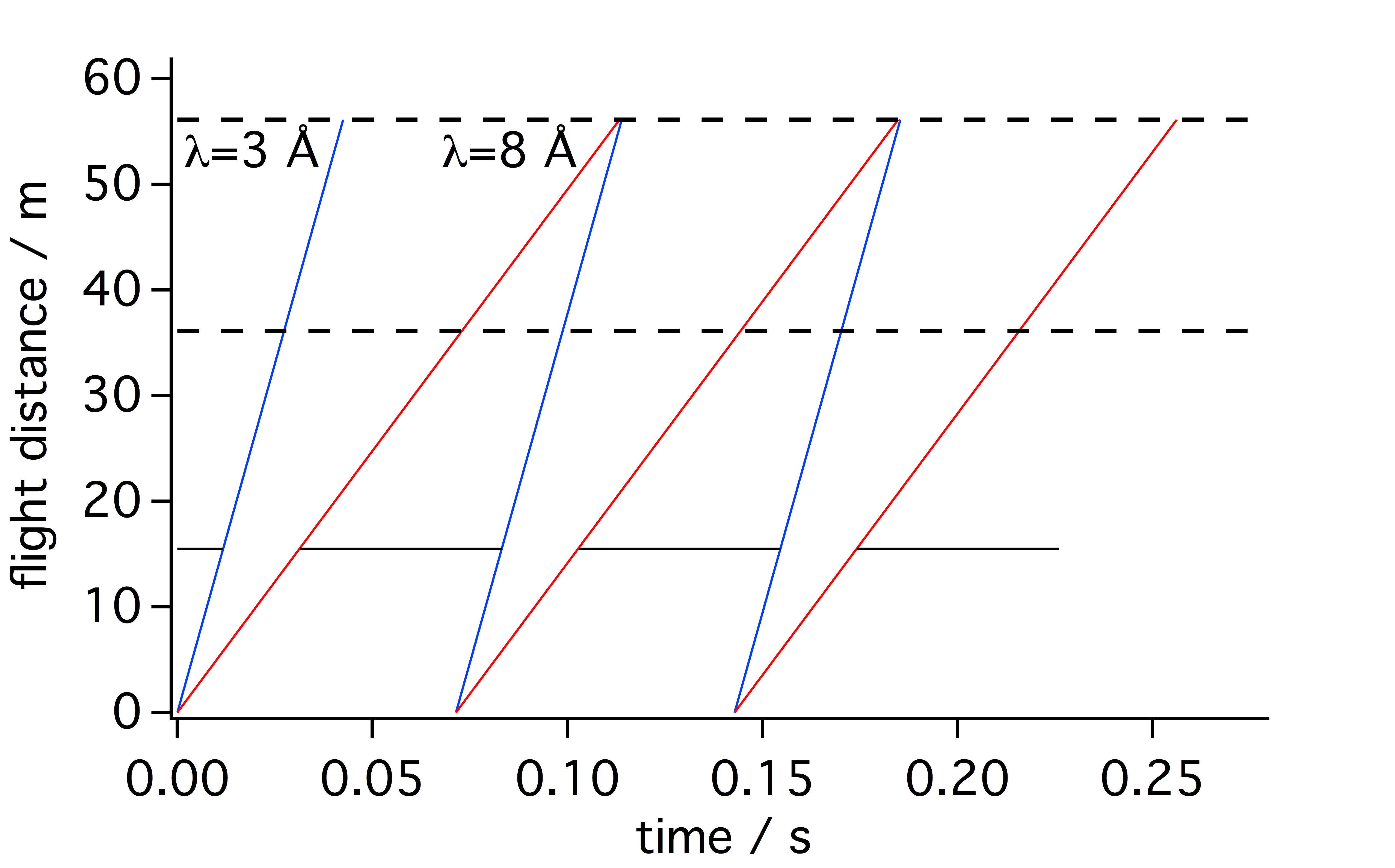}
\end{center}
\caption{Setup with a single frame overlap chopper at 15.5 m. The dashed lines mark the positions of the two detector positions in SKADI. Wavelength of the limiting wavelengths for the wavelengths band are given in the figure.}
\label{fig:justframeoverlap}
\end{figure}

In order to gain a good frame separation between single pulses thus the wavelength band is chosen to be 5 \AA\, wide. Considering the first chopper that can be placed at SKADI is as 15.5 m this results in a configuration as shown in fig. \ref{fig:justframeoverlap}.

\subsection{Suppression of higher orders}
\label{sec:higherOrder}
The configuration as shown in fig. \ref{fig:justframeoverlap} is designed only to transmit neutrons between 3 and 8 \AA\, wavelength. However the velocity distribution in each pulse follows a Maxwell distribution, therefore also neutron with a very long wavelengths are available that can potentially pass in one of the next openings of the chopper. These neutrons are termed higher order neutrons, and depending on whether they are in the second or n-th opening of the chopper they are referred to as neutrons of the second or n-th order. Their upper and lower wavelength limit and the wavelength of the initial pulse are related via

\begin{align}
\lambda _{n-th\,order,fast} & = & \frac{C}{v_{n-th\,order, fast}} = \frac{C\cdot \left(\frac{n-1}{f}+\frac{x_{chopper}}{v_{fast}}\right)}{x_{chopper}},\\
\lambda _{n-th\,order,slow} & = & \frac{C}{v_{n-th\,order, slow}} = \frac{C\cdot \left(\frac{n-1}{f}+\frac{x_{chopper}}{v_{slow}}\right)}{x_{chopper}}.
\end{align}

\begin{figure}[ht]
\begin{center}
\includegraphics[width=0.8\textwidth]{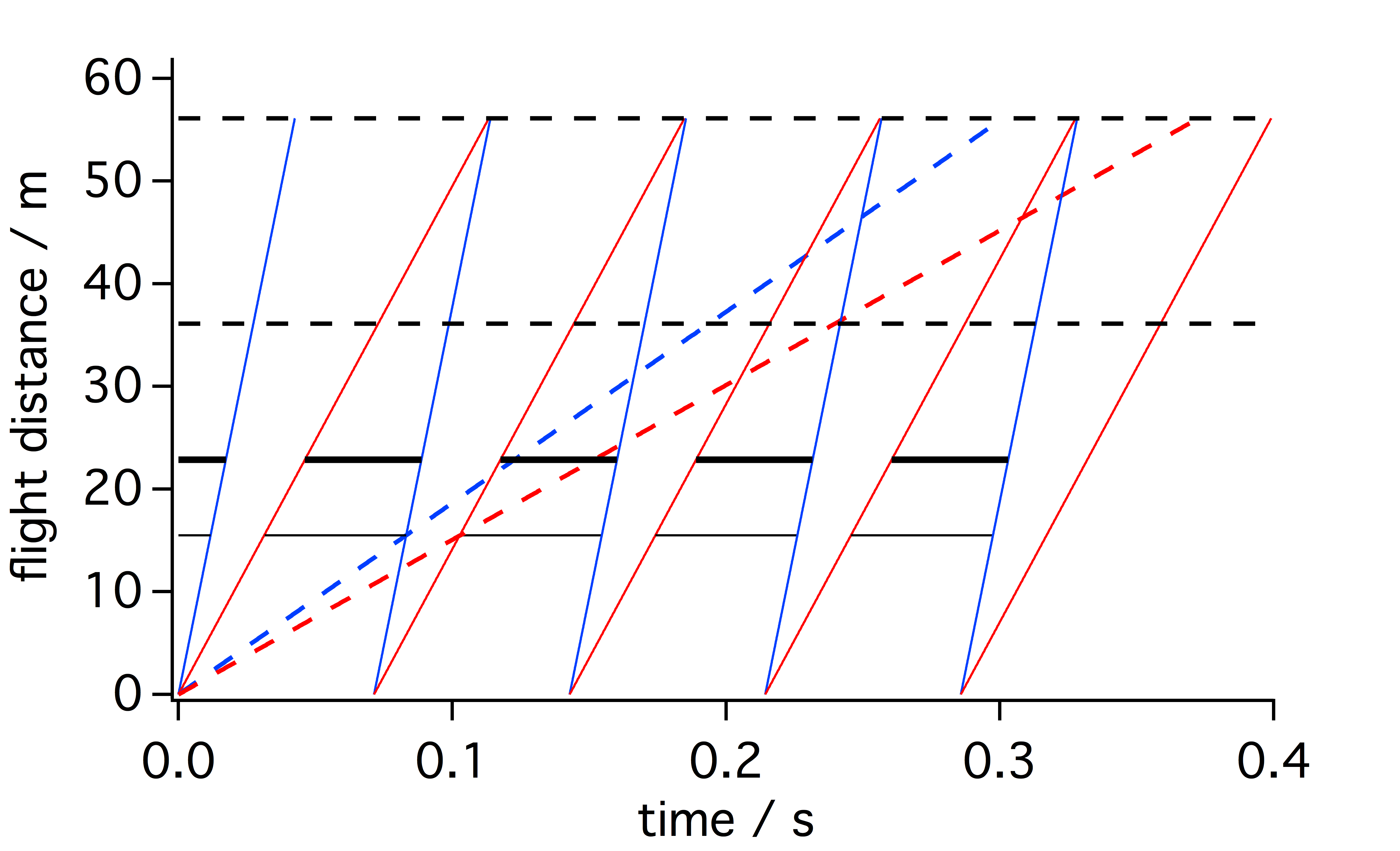}
\end{center}
\caption{Setup with a single frame overlap chopper at 15.5 m and a chopper at 22.85 m for the suppression of higher order neutrons. The dashed lines mark the positions of the two detector positions in SKADI. Wavelength of the limiting wavelengths for the wavelengths band are given in the figure. The second order neutrons pass between the second and third pulse of the neutrons of the first order. Dashed colored lines indicate the path of the fast and slow neutrons of the second order.}
\label{fig:suppresshighorders}
\end{figure}

Here the index n-th order signifies the property relates to the n-th order, $f$ is the frequency of the chopper, $x_{chopper}$ is the chopper distance from the source and $v_{fast}$ and $v_{slow}$ are the fastest and slowest neutrons accepted in the first pulse respectively. Using the numbers given above this renders for the second order $\lambda_{2nd,fast}=21.25$\AA\, and $\lambda_{2nd,slow}=26.25$\AA\,. These orders have to be suppressed by an additional chopper at a later position than the first chopper. An additional requirement is that this second chopper should not cut off neutrons from the primary pulse, i.e. it can only take effect when the slowest neutrons of the next pulse overtake the fast neutrons of the second order pulse as shown in fig. \ref{fig:suppresshighorders}.

The time of this encounter is described by the following equation:

\begin{equation}
t_{encounter}=T\frac{v_{2nd\,order, fast}}{v_{slow}-v_{2nd\,order, fast}}.
\end{equation}

Here $v_{2nd\,order, fast}$ is the velocity of the fastest neutrons in the second order. For the encounter between the slowest neutrons in the second pulse and the fastest neutron of the second order this equation yields $t_{encounter}=0.04313\,$s, which in turn positions the second chopper at

\begin{equation}
x_{chopper,2} = v_{slow}\times t_{encounter}=495\,\mbox{m s}^{-1}\times0.04313\,\mbox{s}=21.35\,\mbox{m}.
\end{equation}

A similar argument can be made for the fast neutrons in the third pulse and the slowest neutrons of the second order, resulting in

\begin{equation}
t_{encounter}=2T\frac{v_{2nd\,order,slow}}{v_{fast}-v_{2nd\,order,slow}}.
\end{equation}

Using the numbers as above this renders $t_{encounter}=0.0184\,$s 

\begin{equation}
x_{chopper,2} = v_{fast}\times t_{encounter}=1320\,\mbox{m s}^{-1}\times0.0184\,\mbox{s}=24.28\,\mbox{m}.
\end{equation}

Between these two positions the second chopper for the suppression of higher orders can be moved with impunity. To maximise overlap protection in both directions it should be placed central between these two positions, i.e. for example at $x_{chopper,2}=22.85\,$m as depicted in fig. \ref{fig:suppresshighorders}.

Up to now only second order neutrons were considered. However higher order neutrons could appear with very low wavelength and albeit low intensity. These will however be suppressed in the collimation. The distance $x_{fall}$ a neutron of a given wavelength falls down due to gravitation over a given collimation distance $s_{coll}$ is given by

\begin{equation}
x_{fall}=\frac{1}{2}g\left(\frac{s_{coll}\lambda}{C}\right)^2.
\end{equation}
Here $g$ is the gravitational acceleration. If $x_{fall}$ is larger than the maximum collimation opening the neutron will be lost in the collimation in all cases. For SKADI the maximum collimation opening is $x_{collOpen}=0.03\,$m.  We can use this limit to find the upper limit for the wavelength of neutrons that can be transmitted through the collimation with

\begin{equation}
\lambda \leq \sqrt{\frac{2 x_{collOpen}}{g}\left(\frac{C}{s_{coll}}\right)^2}
\end{equation}

For collimation distances as used in SKADI, where $s_{coll}$= 8, 14 and 20 m, this renders $\lambda=$38.7, 22.1 and 15.5 \AA\, respectively. Comparing this with the wave length limits for the higher order neutrons from above it can be seen that only second order neutrons pass the collimation as the fastest neutrons of the third order have a wavelength of 39.5 \AA, hence only the second order neutrons need to be suppressed with an additional chopper.

\subsection{Penumbra}
\label{sec:penumbra}
\begin{figure}[ht]
\begin{center}
\includegraphics[width=0.8\textwidth]{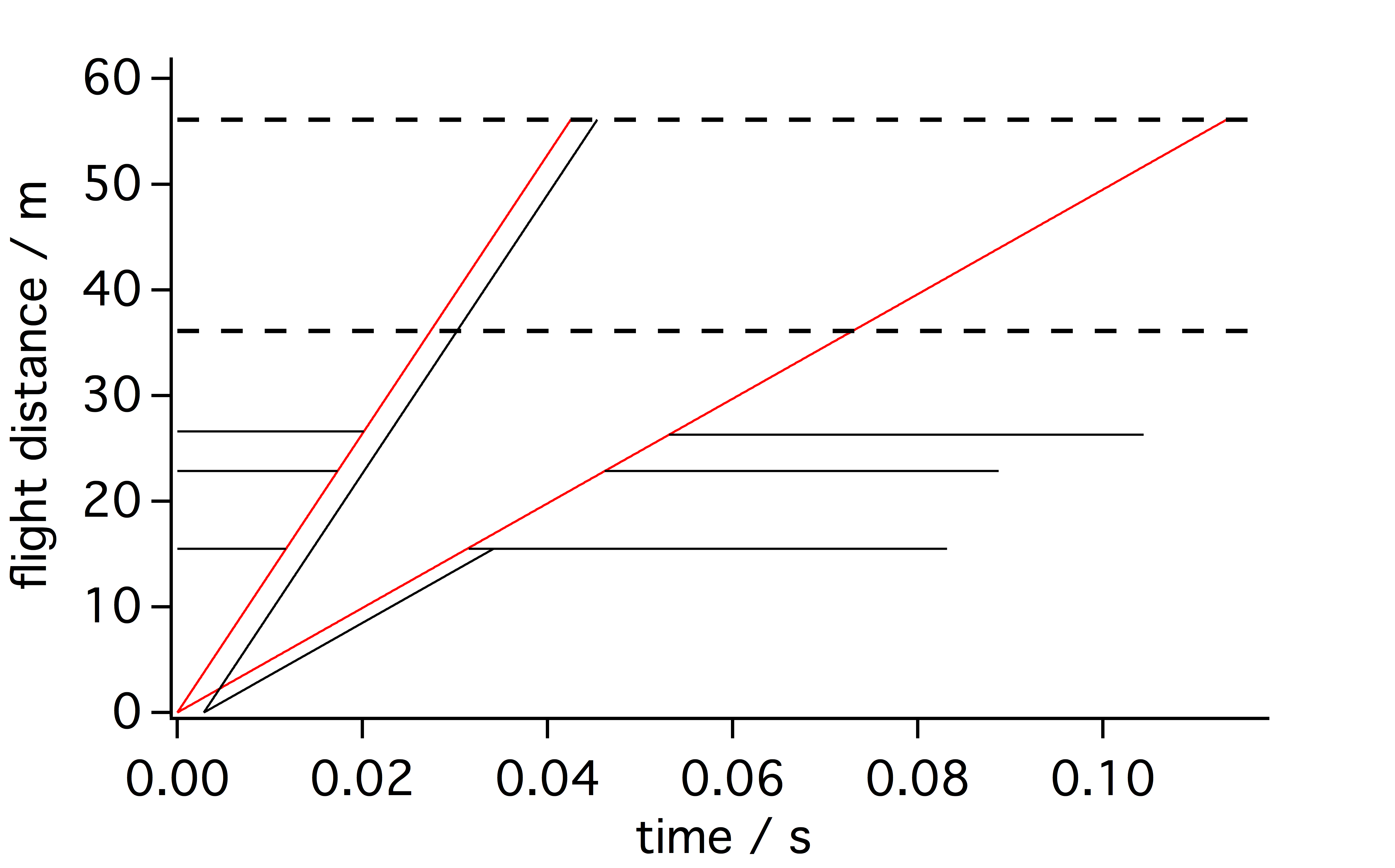}
\end{center}
\caption{Description of penumbra. Black lines indicate the parts of the spectrum that are emitted at the beginning or end of the pulse and hence create a penumbra at the detector.}
\label{fig:penumbra}
\end{figure}

As the pulse of a spallation source has a finite length and neutrons of the whole wave length band are produced during that pulse at the edges of the wavelength pulse at the detector position there is a penumbra of either neutrons slower than desired that are created at the very beginning of the pulse or faster than desired and created at the very end of the pulse. The situation is depicted in fig. \ref{fig:penumbra}.

With a pulse length of $t_{pulse}$ the velocity of this neutrons that can still pass a chopper is given by

\begin{align}
v_{too\,slow} &=& \frac{x_{chopper}}{\frac{x_{chopper}}{v_{slow}}+t_{pulse}},\\
\label{eq:penumbra1}
v_{too\,fast} &=& \frac{x_{chopper}}{\frac{x_{chopper}}{v_{fast}}-t_{pulse}}.
\end{align}

For the calculations as shown above this yields the values given in table \ref{tab:penumbra}.

\begin{table}
\begin{center}
\begin{tabular}{c|c|c|c|c}

$x_{chopper}$/m & $v_{too\,slow}$/m s$^-1$ & $\lambda_{too\,slow}$/\AA & $v_{too\,fast}$/m s$^-1$ & $\lambda_{too\,fast}$/\AA \\ 
\hline 
15.5  & 452.2 & 8.6 & 1733.3 & 2.3 \\ 
\hline 
22.85 & 464.4 & 8.5 & 1574.7 & 2.5 \\ 

\end{tabular}
\end{center}
\caption{Too fast or too slow neutrons that may slip by the choppers, due to a finite width of the pulse length.}
\label{tab:penumbra} 
\end{table}

Only a chopper located directly at the detector plane could remove at least the too slow neutrons. The too fast neutrons would then still be contained in the spectrum of the slower neutrons. Thus adding an additional chopper pair further downstream at the instrument improves the penumbra but can ultimately not remove it, so it is an inherent contribution to the wavelength distribution of the instrument.

\section{High resolution option}
\begin{figure}[ht]
\begin{center}
\includegraphics[width=0.8\textwidth]{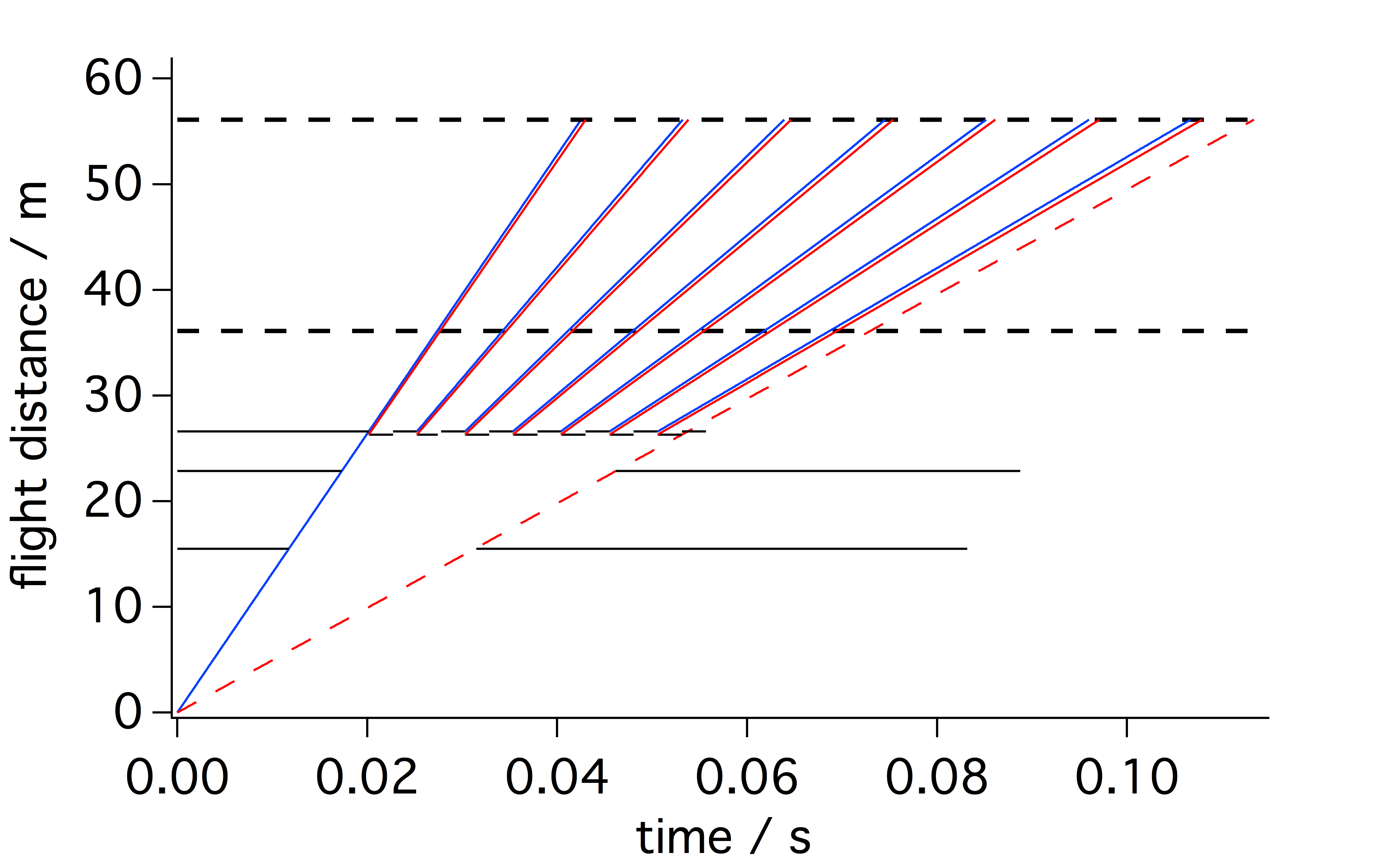}
\end{center}
\caption{Setup with high resolution choppers for $\Delta\lambda/\lambda=0.01$. The dashed lines mark the positions of the two detector positions in SKADI. Wavelength of the limiting wavelengths for the wavelengths band are given in the figure. Each single pulse has a distinct wavelength distribution (see text).}
\label{fig:oneperchop}
\end{figure}

In order to gain a low $\Delta\lambda/\lambda$ resolution of the wavelength a system as described by Dewhurst et al. can be employed \cite{Dewhurst}. Alternatively, also crystal monochromators \cite{Mildner2001} can be used, which generally show better resolutions than velocity selectors \cite{Feoktystov}. As a chopper systems will be in place at SKADI anyway, using the first approach seems more sensible. A quite practical advantage of the first solution is also, that a constant $\Delta\lambda/\lambda$ can be reached for all wavelengths. In order to achieve this, two single disk choppers are offset by a distance $\Delta D$ the rear Chopper opens at the same time when the front chopper closes. The details of the setup can be seen in fig.\ref{fig:oneperchop}.

In contrast to the solution described in \cite{Dewhurst} the initial pulse at a spallation source is fixed in frequency and cannot be adapted to shift to accomodate a nearly continuous relation between transmission and resolution. Regarding the relation 

\begin{equation}
\Delta \lambda/\lambda = \Delta D / D
\end{equation}

with $\Delta D$ being the distance between two resolution shoppers and $D$ being the distance between the center between these two choppers and the detectors it would be optimal to use extremely close choppers very far away from the detector. However, as described in section \ref{sec:penumbra}, the overall performance of the instrument can benefit from choppers further downstream from the frame overlap choppers. In the case of SKADI the last 8 m before the sample position are occupied by a set of optics for higher resolution, which also restricts the positioning of the high resolution choppers in that direction. The last possibility to position a chopper at SKADI is at 26.6 m. In order to achieve a resolution of $\Delta D/D = a$ with two choppers at positions $x_{chopper,1}$ and $x_{chopper,2}$ with a detector at position $x_D$ the relation between the distances is given by

\begin{equation}
x_{chopper,2} = -\frac{1}{1-\frac{a}{2}}\left(a x_D - x_{chopper,1}\left[1+\frac{a}{2}\right]\right).
\end{equation}

Using $x_D=56.11\,$m and $x_{chopper,1}=26.6\,$m and $a=0.01$ this yields $x_{chopper,2}=26.3\,$m. In order to operate also in the flux optimized mode these choppers will need to open and close for the same neutrons as accepted by the previous chopper setup, i.e. from $\lambda=3$ to $\lambda=8\,$\AA.

The width of the transmitted wavelength band is given by

\begin{equation}
\lambda_{fast}-\lambda_{slow}=C t\left(\frac{1}{x_{chopper,1}}-\frac{1}{x_{chopper,2}}\right).
\end{equation}

As this is strictly linear in $t$ and $t$ is given by the time of arrival at the first chopper, the widest wavelength band is given by the slowest neutrons at $\lambda=8\,$\AA\, with $t=0.05341\,$s. This results in wavelength band of $\lambda_{fast}-\lambda_{slow}=8.96\cdot10^{-2}\,$\AA. For $\lambda=3\,$\AA\, the values are $t=0.01992\,$s and $\lambda_{fast}-\lambda_{slow}=3.38\cdot10^{-2}\,$\AA. At the detector positions the time differentials between the fastest and the slowest neutrons in each high resolution neutron pulse is given by

\begin{equation}
t_{fast}-t_{slow}=\frac{x_D}{C}\left(\lambda_{fast}-\lambda_{slow}\right).
\end{equation}

With the values as given before this results in $t_{fast}-t_{slow}=1.3\cdot10^{-3}\,$s for $\lambda=8\,$\AA\, and $t_{fast}-t_{slow}=4.8\cdot10^{-4}\,$s for $\lambda=3\,$\AA.

Given the fact that only integer values of the base frequency of the source can be used it is sensible to calculate the limiting frequency, after which frame overlap for the high resolution secondary pulses occur, with the the larger of these two values.

The limiting frequency is given by

\begin{equation}
T>N\times \left(t_{fast}-t_{slow}\right).
\end{equation}
where $N$ is an integer multiple of the base frequency of the source. Resolving this leads to $N=54$ with the values as given above, which would allow 53 high resolution pulses to be transmitted during each base frequency pulse of the source. For the choppers this would mean a frequency of 756 Hz which is not practically relevant for a SANS instrument such as SKADI as these high speeds could only achieved with a specialized chopper setup. At $f=210\,$Hz and $N=15$ fourteen high resolution pulses could be transmitted for every base frequency pulse of the source. Here it becomes visible that for the high resolution option the flux is directly linear to the frequency of the high resolution choppers as stated by Dewhurst in \cite{Dewhurst}.

\section{Resolution}
\label{sec:resolution}
Resolution in this manuscript only focuses on the wavelength resolution $\Delta \lambda/\lambda$ unless stated otherwise.

At a spallation source there are several distinct properties on which the resolution depends:
\begin{itemize}
\item The length of a single pulse at the source $t_{pulse}$
\item The length of the instrument $D$
\item The time a chopper needs for closing/opening when traversing the cross section of the neutron guide $t_{traverse}$.
\end{itemize}

In the following the influence of each of these properties will be discussed. The resolution as discussed here only is applicable in the high-flux mode of the setup as described above, as the high-resolution mode yields a constant resolution of $\Delta\lambda/\lambda=0.01$.

\subsection{Length of the Pulse $t_{pulse}$ and Length of the Instrument $D$}
As the detector is only time sensitive and not wavelength sensitive it is impossible to determine at what time in the pulse a neutron was emitted. Thus, at all times at the detector position the best achievable time resolution for any given neutron is

\begin{equation}
\frac{t_{pulse}}{t_{full\,Distance}} = \frac{t_{pulse} v_{neutron}}{D}=\frac{t_{pulse} C}{D \lambda}.
\end{equation}

\begin{figure}[ht]
\begin{center}
\includegraphics[width=0.8\textwidth]{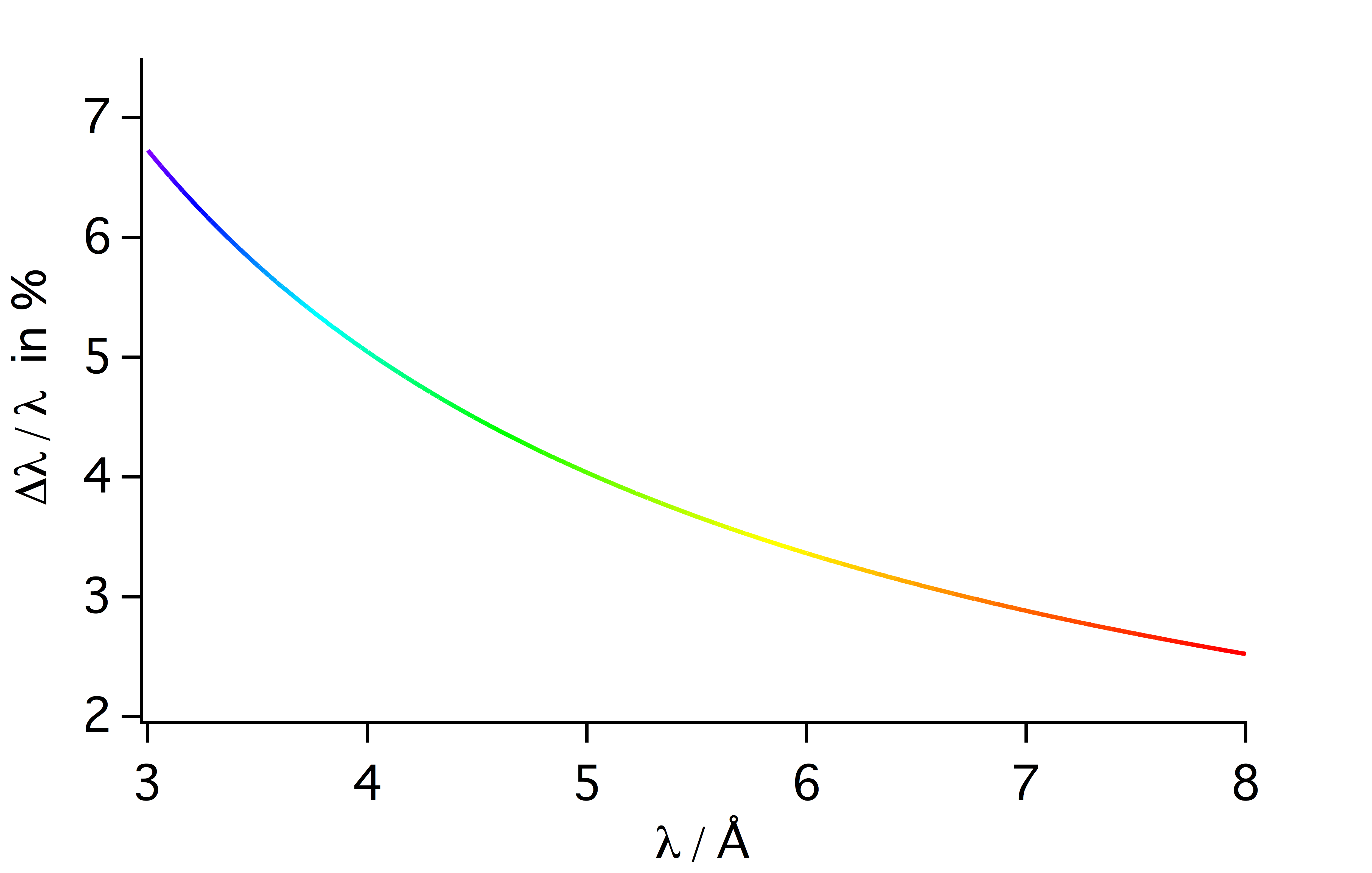}
\end{center}
\caption{Dependency of wavelength resolution depending on wavelength as a result to finite pulse width.}
\label{fig:resolutionpulselength}
\end{figure}

Here ${t_{full\,Distance}}$ is the time for a neutron to travel from the moderator to the detector and $v_{neutron}$ is the velocity of that neutron with the wavelength $\lambda$. A plot of the wavelength dependency of the resolution is given in fig. \ref{fig:resolutionpulselength}. It can be seen that the resolution due to the finite width of the pulse is in no case worse than 7\% in the case of SKADI.

\subsection{Opening and Closing of Choppers $t_{traverse}$}
During the time it takes a chopper to close the full cross section of the neutron guide neutrons of other wavelength can still pass the chopper. However, they also need to be close enough to the chopper for that to occur. The closing time $t_{traverse}$ is given by

\begin{equation}
t_{traverse} = \frac{x_{NG}}{2\pi f r}.
\label{eq:tTrav}
\end{equation}

Here $x_{NG}$ is the width of the neutron guide and $r$ the radius of the chopper disc where is intersects with the neutron guide the farthest away from the rotational axis of the chopper. Using this time is can be used to calculate the wavelength spread of the transmitted neutrons with

\begin{equation}
\Delta \lambda = \frac{\Delta \lambda}{d t}\Delta t=\frac{C}{s}\cdot \Delta t=\frac{\lambda}{t_{time\,to\,chopper}}\Delta t=\lambda\frac{t_{traverse}}{t_{time\,to\,chopper}} 
\end{equation}

Here s is the distance between the neutron and the chopper when the chopper starts to close. It also should be noted that $t_{traverse}$ is identified with $\Delta t$. Neutrons of different wavelength need different times to reach the chopper, hence we get

\begin{equation}
\frac{\Delta \lambda}{\lambda} = \frac{s \lambda}{C}\Delta t
\end{equation}

\begin{figure}[ht]
\begin{center}
\includegraphics[width=0.8\textwidth]{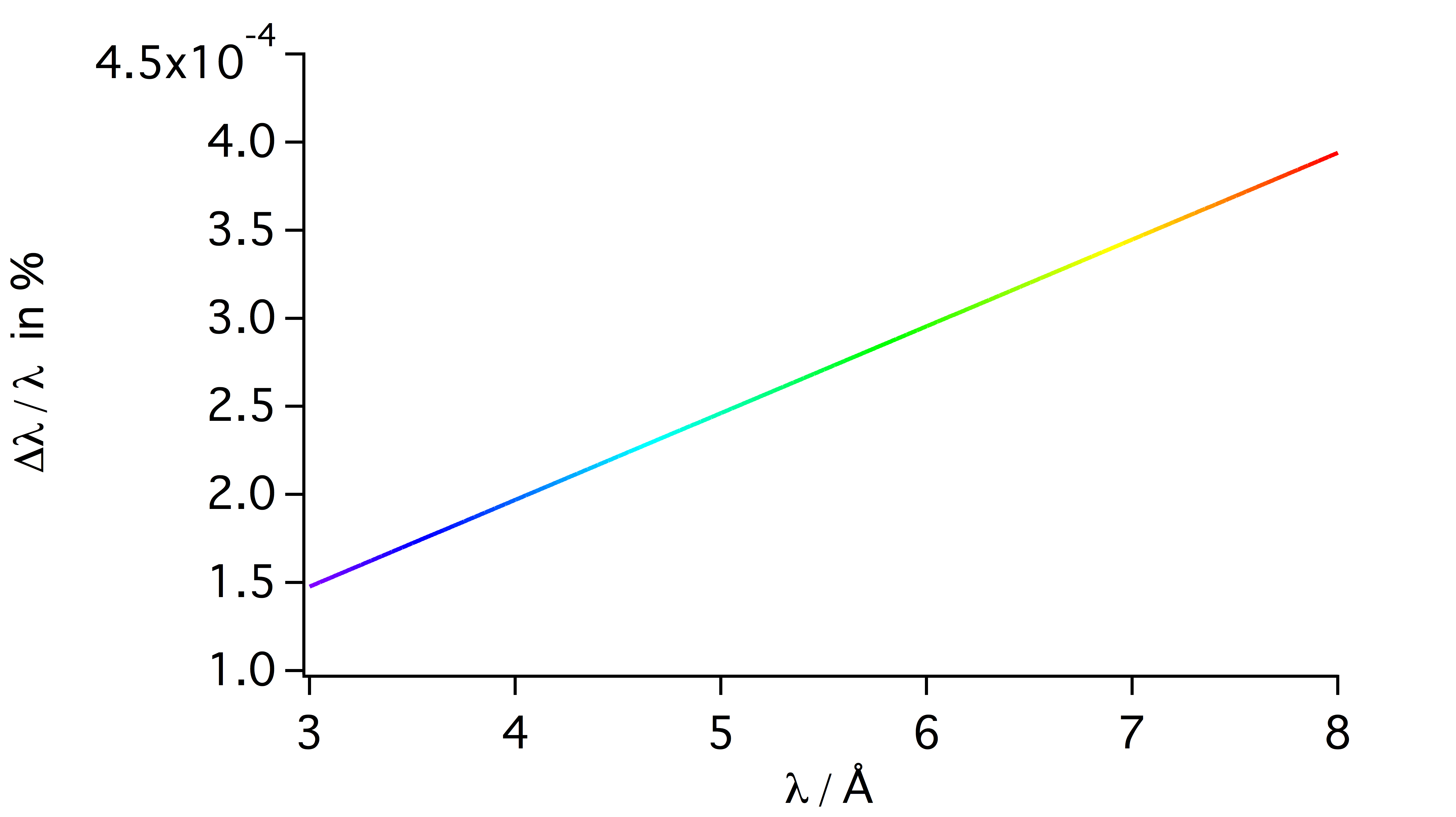}
\end{center}
\caption{Dependency of wavelength resolution depending on wavelength as a result of finite closing time of the choppers. }
\label{fig:resolutionchopperopening}
\end{figure}

A plot with of this function can be seen in fig. \ref{fig:resolutionchopperopening}. The longest distance for $s=1.601\,$m, which results from the fastest neutrons at $\lambda=3\,$\AA. From the plot it can be seen that the values for the error are around $\Delta \lambda=1\cdot 10^{-14}$, which results in values of $\Delta \lambda /\lambda \sim 10^{-4}$\%, which for most applications can be neglected.

%
%
%
%
%

\section{Summary and Conclusion}
Considerations about the positioning of choppers, modes and frequencies of operation and the impact on the resolution were presented.

The resulting set of choppers comprises a frame overlap chopper at 15.5 m, a chopper for the suppression of higher orders at 22.85 m and a high-resolution chopper pair at 26.3 and 26.6 m using the dimensions and restrictions as they are the case for the SKADI SANS instrument at the upcoming ESS.

Two modes of operation are possible, where one is yielding the highest possible flux with a wavelength dependent resolution and the other is yielding a constant resolution of $\Delta\lambda/\lambda=0.01$ with a reduced flux. This flux is linear with the frequency the high resolution chopper can achieve.

This chopper setup fulfills all the requirements on a chopper set at a SANS instrument, namely definition of wavelength band, optimization of flux and/or resolution and providing a clear neutron spectrum. Using this setup in SKADI at the upcoming ESS it will enable users at SKADI to perform high profile experiments and allows them to tune the performance of SKADI as is best suited for their experiment. This emphasizes the role of SKADI as a versatile SANS instrument, making it complementary with the other SANS instrument LoKi \cite{Jackson2013} at the ESS and thus helping to a achieve a complementary neutron instrument suite, which is desireable to cover a wide range of experimental needs of scientific users.

\section{Acknowledgements}
I gratefully acknowledge discussions with my colleagues Henrich Frielinghaus, Charles Dewhurst, Nikolaos Tsapatsaris, Sylvain D\'{e}sert and Patrice Permingeat and thank for the inspirations drawn from those discussions.

\bibliography{literature}

\begin{thebibliography}{1}

\bibitem{Jackson2013}
Kalliopi~Kanaki Andrew~Jackson.
\newblock Loki - a broadband sans instrument, 2013.

\bibitem{Dewhurst}
C.~D. Dewhurst, I.~Grillo, D.~Honecker, M.~Bonnaud, M.~Jacques, C.~Amrouni,
  A.~Perillo-Marcone, G.~Manzin, and R.~Cubitt.
\newblock {The small-angle neutron scattering instrument D33 at the Institut
  Laue{--}Langevin}.
\newblock {\em Journal of Applied Crystallography}, 49(1):1--14, Feb 2016.

\bibitem{Feoktystov}
Artem~V. Feoktystov, Henrich Frielinghaus, Zhenyu Di, Sebastian Jaksch, Vitaliy
  Pipich, Marie-Sousai Appavou, Earl Babcock, Romuald Hanslik, Ralf Engels,
  G{\"{u}}nther Kemmerling, Harald Kleines, Alexander Ioffe, Dieter Richter,
  and Thomas Br{\"{u}}ckel.
\newblock {KWS-1 high-resolution small-angle neutron scattering instrument at
  JCNS: current state}.
\newblock {\em Journal of Applied Crystallography}, 48(1):61--70, Feb 2015.

\bibitem{jaksch2014}
S.~Jaksch, D.~Martin-Rodriguez, A.~Ostermann, J.~Jestin, S.~Duarte Pinto, W.G.
  Bouwman, J.~Uher, R.~Engels, and H.~Frielinghaus.
\newblock Concept for a time-of-flight small angle neutron scattering
  instrument at the european spallation source.
\newblock {\em Nuclear Instruments and Methods in Physics Research Section A:
  Accelerators, Spectrometers, Detectors and Associated Equipment}, 762:22 --
  30, 2014.

\bibitem{Mildner2001}
D.~F.~R Mildner, M.~Arif, and S.~A. Werner.
\newblock {Neutron transmission through pyrolytic graphite monochromators}.
\newblock {\em Journal of Applied Crystallography}, 34(3):258--262, Jun 2001.

\bibitem{HALLWILTON2013}
Camille Theroine, Guillaume Pignol, Torsten Soldner, Richard Hall-Wilton, and
  Camille Theroine.
\newblock Ess science symposium on neutron particle physics at long pulse
  spallation sources, nppatlps 2013 status of the european spallation source
  ess ab, the instrument selection process, and a fundamental physics beamline
  at the ess.
\newblock {\em Physics Procedia}, 51:8 -- 12, 2014.

\end{thebibliography}
\bibliographystyle{plain}

\end{document}